# Measurements and semi-empirical calculations of $CO_2$+$CH_4$ and $CO_2$+$H_2$ collision-induced absorption across a wide range of wavelengths and temperatures. Application for the prediction of early Mars surface temperature.

Martin Turbet[a,#], Christian Boulet[b] and Tijs Karman[c]

[a] Observatoire astronomique de l'Université de Genève, 51 chemin des Maillettes, 1290 Sauverny, Switzerland
[b] Université Paris-Saclay, CNRS, Institut des Sciences Moléculaires d'Orsay (ISMO), 91405, Orsay, France.
[c] Harvard-Smithsonian Center for Astrophysics, Atomic and Molecular Physics Division, Cambridge, MA, USA.

[#] Corresponding author: martin.turbet@unige.ch

**Abstract**

Reducing atmospheres have recently emerged as a promising scenario to warm the surface of early Mars enough to drive the formation of valley networks and other ancient aqueous features that have been detected so far on the surface of Mars. Here we present a series of experiments and calculations to better constrain $CO_2$+$CH_4$ and $CO_2$+$H_2$ collision-induced absorptions (CIAs) as well as their effect on the prediction of early Mars surface temperature. First, we carried out a new set of experimental measurements (using the AILES line of the SOLEIL synchrotron) of both $CO_2$+$CH_4$ and $CO_2$+$H_2$ CIAs. These measurements confirm the previous results of Turbet et al. 2019, Icarus vol. 321, while significantly reducing the experimental uncertainties. Secondly, we fitted a semi-empirical model to these CIAs measurements, allowing us to compute the $CO_2$+$CH_4$ and $CO_2$+$H_2$ CIAs across a broad spectral domain (0-1500cm$^{-1}$) and for a wide range of temperatures (100-600K). Last, we performed 1-D numerical radiative-convective climate calculations (using the LMD Generic Model) to compute the surface temperature expected on the surface of early Mars for several $CO_2$, $CH_4$ and $H_2$ atmospheric contents, taking into account the radiative effect of these revised CIAs. These calculations demonstrate that thick $CO_2$+$H_2$-dominated atmospheres remain a viable solution for warming the surface of Mars above the melting point of water, but not $CO_2$+$CH_4$-dominated atmospheres. Our calculated $CO_2$+$CH_4$ and $CO_2$+$H_2$ CIA spectra and predicted early Mars surface temperatures are provided to the community for future uses.

## 1. Introduction

Understanding how the early Martian climate could have been warm enough for liquid water to flow on the surface is still one of the major enigmas of planetary science (Wordsworth 2016, Haberle et al., 2017, Ramirez & Craddock 2018). Reducing atmospheres have recently emerged as one of the most promising scenarios to solve this enigma (Ramirez et al., 2014, Wordsworth et al., 2017). Specifically, it has been proposed that significant amounts of hydrogen and methane could have accumulated in the atmosphere of early Mars 3-4 billion years ago through (i) volcanic outgassing from a reduced mantle (Ramirez et al., 2014), (ii) serpentinization (Chassefière et al., 2016), (iii) radiolysis (Tarnas et al., 2018), (iv) clathrate release e.g. after atmospheric collapse and reinflation events (Kite et al., 2020) or (v) impact-induced thermochemistry (Haberle et al., 2019).

The community's recent interest in this scenario stems mostly from the fact that $H_2$ and $CH_4$ are the only known yet plausible gases that could theoretically raise the surface temperature of a $CO_2$-dominated early Martian atmosphere above the melting point of water, i.e. the typical temperature required to form aqueous mineralogic and geomorphologic features visible today on the surface of Mars.

It has been shown that reducing gases $CH_4$ and $H_2$ can in fact produce a strong greenhouse effect in a $CO_2$-dominated atmosphere via their far-infrared absorptions (CIAs) induced by collisions with $CO_2$ (Ramirez et al., 2014, Wordsworth et al., 2017). Initially, the modelling of these CIAs was based, due to lack of relevant data, on the CIAs of $H_2$-$N_2$ and $CH_4$-$N_2$ pairs, respectively (Ramirez et al., 2014).

Since the initial work of Ramirez et al. (2014), significant efforts have been made to better constrain the intensity and spectral shape of $CO_2$+$CH_4$ and $CO_2$+$H_2$ CIAs. First, Wordsworth et al. (2017) provided theoretical calculations of these CIAs, using a semi-empirical model that assumes that the shape of the spectrum of $CO_2$+$H_2$ ($CO_2$+$CH_4$, respectively) CIA can be approximated as a linear combination of $CO_2$+$CO_2$ and $H_2$+$H_2$ ($CH_4$+$CH_4$, respectively) CIAs, which are known (Richard et al., 2012, Karman et al., 2019). The relative contribution of the two pure gas $CO_2$+$CO_2$ and $H_2$+$H_2$ ($CH_4$+$CH_4$, respectively) CIAs was taken equal to what is necessary to fit $N_2$+$H_2$ ($N_2$+$CH_4$, respectively) for which measurements have been performed, with $N_2$+$N_2$ and $H_2$+$H_2$ ($CH_4$+$CH_4$, respectively) pure gas CIAs. The integrated $CO_2$+$H_2$ ($CO_2$+$CH_4$, respectively) CIA spectrally integrated intensity was then predicted based on ab initio calculations of the zeroth order spectral moment (Wordsworth et al. 2017). With this model, Wordsworth et al. (2017) showed that the $H_2$-$CO_2$ and $CH_4$-$CO_2$ CIAs are more intense than the $H_2$-$N_2$ and $CH_4$-$N_2$ CIAs, respectively, mostly as a result of the fact that $CO_2$ has larger polarizability and multipole moments than $N_2$.

Turbet et al. (2019) then performed the first measurements of those $H_2$-$CO_2$ and $CH_4$-$CO_2$ CIAs. They showed that Wordsworth et al. (2017) had likely overestimated their average contribution, possibly by a factor of 1.6-1.7, at room temperature, and in the 50-550cm$^{-1}$ spectral range.

Knowing precisely the value of these CIAs is crucial to evaluate the true greenhouse effect of reducing atmospheres, and more specifically to estimate the minimum amount of $H_2$ (or $CH_4$, respectively) in a thick $CO_2$-dominated atmosphere required to warm the surface of early Mars above the melting point of water. This calculation is critical to assess the credibility of the reducing early Mars atmosphere scenario. On the one hand, maintaining hydrogen concentrations above a few percent in the early Mars atmosphere would be difficult, given the predicted escape rates of hydrogen to space (Batalha et al., 2015). On the other hand, accumulating methane concentration in excess of 10% in the early Mars atmosphere should lead to the formation of photochemical hazes (Trainer et al., 2006), leading to a significant cooling of the surface. With the CIAs calculated in Wordsworth et al. (2017), only a few % of $CH_4$ and/or $H_2$ in a $CO_2$-dominated atmosphere could suffice to warm early Mars enough for surface liquid water to become stable (Wordsworth et al., 2017, Ramirez, 2017).

Here we performed a series of experiments and calculations to better constrain $CO_2$+$CH_4$ and $CO_2$+$H_2$ collision-induced absorption spectra as well as to determine their impact on the prediction of early Mars surface temperature. We present in Section 2 the results of our latest measurements of the $CO_2$+$H_2$ and $CO_2$+$CH_4$ CIAs. In Section 3, we present our semi-empirical model and show the results of our calculated CIAs. In Section 4, we use a numerical climate model to estimate the effect of these CIAs on surface temperature of early Mars. In Section 5, we present our conclusions as well as discuss future pathways to improve these calculations.

## 2. Updated experimental CIA measurements

After the first measurement campaign of $CO_2$+$H_2$ and $CO_2$+$CH_4$ CIAs carried in April 2018 and whose results were presented in Turbet et al. (2019), a second set of experiments was carried in September 2018. A total of fourteen $CO_2$+$H_2$ infrared spectra (14 spectra between 301 and 942 mbar, for a mixture of ∼ 50% $CO_2$ and 50% $H_2$) and thirteen $CO_2$+$CH_4$ infrared spectra (13 spectra between 301 and 930 mbar, for a mixture of ∼ 50% $CO_2$ and 50% $CH_4$) were recorded during one week, using the experimental setup (AILES line, at the SOLEIL synchrotron facility) detailed in Section 2.1 of Turbet et al. (2019). These spectra were then post-treated and converted into two collision-induced absorption (CIA) spectra (see Fig. 1; green spectra) using the methodology described in Section 2.2 of Turbet et al. (2019).

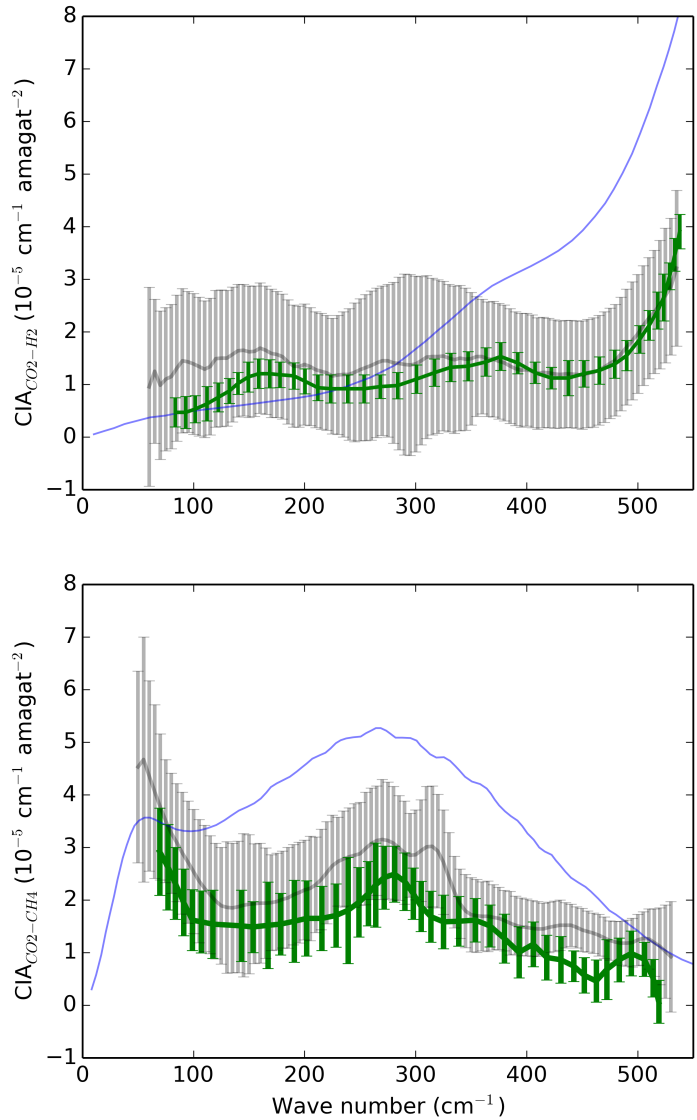

*Figure 1:* In green, $H_2$+$CO_2$ CIA (top panel) and $CH_4$+$CO_2$ CIA (bottom panel) measured in this work at room temperature (symbols with error bars) in September 2018 measurement campaign. In black, the previously measured (Turbet et al., 2019) $H_2$+$CO_2$ CIA (top panel) and $CH_4$+$CO_2$ CIA (bottom panel), also at room temperature. In blue, the previously calculated CIAs from Wordsworth et al. (2017).

Fig. 1 shows the resulting experimental CIA spectra (green spectra), obtained at room temperature (296 K) between ∼ 50 and ∼ 550 $cm^{-1}$. Newly recorded $CO_2$-$H_2$ and $CO_2$-$CH_4$ spectra are both compatible within error bars with the previous measurements (black spectra) of Turbet et al. (2019).

While this gives us more confidence on the reliability of these experimental CIAs, we encourage independent measurements to further test these results (Godin et al., 2019). The uncertainties on the $CO_2+H_2$ CIA measurements (Fig. 1, top panel) have significantly reduced when compared to Turbet et al. (2019). This stems from the fact that particular care was taken to ensure that the Helium-cooled Si-bolometer detector had reached its steady-state temperature of 4.2 K. For our $CO_2+CH_4$ measurements (Fig. 1, bottom panel) as well as for the measurements presented in Figure 7 of Turbet et al. (2019), it turns out that the detector had not fully reached its steady-state equilibrium temperature, thus adding an extra source of uncertainty. The stability of the detector temperature (at 4.2 K) is key to keeping a stable baseline from one experiment to the next (as well as relative to reference spectra), and thus maximizing the signal to noise ratio of the CIA measurements.

The new experimental CIAs presented in Fig. 1 are used for the calibration of the semi-empirical CIA model described in the next Section. The goal of this model is to cover a wider temperature and wavenumber range than in the experiments.

## 3. Predictions of CIAs across a wide range of wavenumbers and temperatures

The most rigorous method for the calculation of CIA spectra is a close-coupling approach which fully takes into account the anisotropy of the interaction potential (see Karman et al., 2015). However, this approach remains hardly tractable with $CO_2$ as a collisional partner, due to the number of collisional channels that must be taken into account and subsequent computer cost.

In this section, we present the results of a tractable but more simple, semi-empirical model designed to calculate CIA spectra in the temperature and spectral ranges of interest for early Mars. Specifically, this model allows to compute the roto-translational band of the CIA spectra of $CO_2$-$H_2$ and $CO_2$-$CH_4$ pairs with the following two objectives:

i   Being in agreement with the available experimental data (see Fig. 1) obtained at room temperature and for a rather limited spectral range.
ii  But giving the possibility to calculate the CIA spectra over all frequencies between 0 and 1500cm$^{-1}$ and at temperatures between 100 and 600K.

Detailed information on this semi-empirical model (construction of the model, validation against experiments) can be found in Appendix A.

Figure 2 shows the results of our calculated $CO_2+H_2$ and $CO_2+CH_4$ CIAs at 200K and 300K between 0 and 1200 cm$^{-1}$, compared with the results of Wordsworth et al. (2017) as well as with the experimental results presented in the previous section (see Fig. 5 and 8 in Appendix A for a more detailed comparison). Our computed CIAs at room temperature match closely with the experimental results presented in Turbet et al. (2019) and in the Section 2 of this paper. The CIAs calculated with our semi-empirical model are also consistently weaker than those predicted in Wordsworth et al. (2017) in the main infrared window of $CO_2$, between 200 and 550cm$^{-1}$. This not only confirms the result of Turbet et al. (2019) but also extends it to a broader range of temperatures.

Our calculated CIAs are provided as ascii files in the Supplementary Materials of this paper. They are used for the numerical climate calculations presented in the next Section. The goal of these climate calculations is to quantitatively evaluate the greenhouse effect of $CO_2+CH_4$ and $CO_2+H_2$ atmospheres on the surface temperature of early Mars.

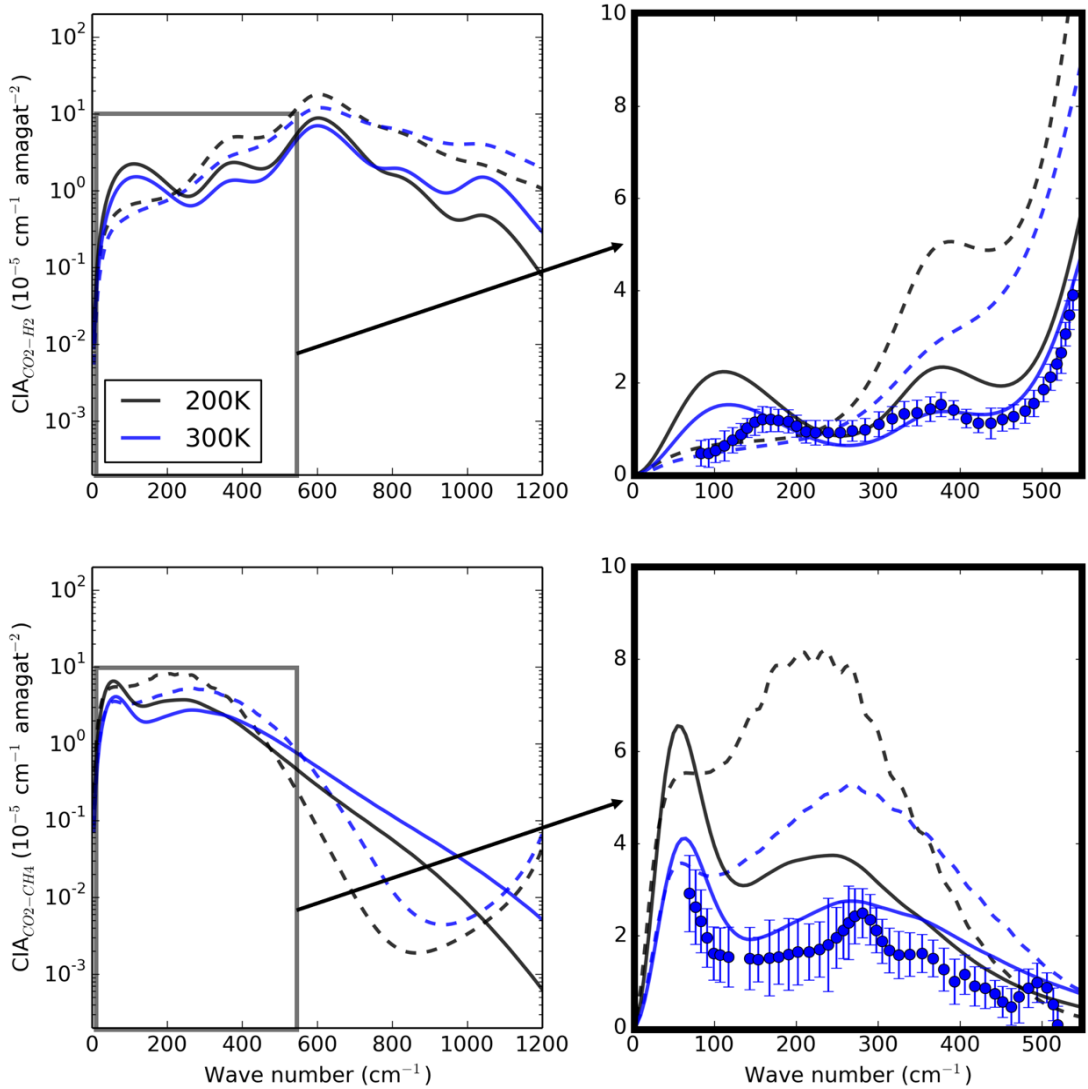

*Figure 2:* $H_2+CO_2$ CIA (top panels) and $CH_4+CO_2$ CIA (bottom panels) calculated in Wordsworth et al. 2017 (dashed lines) and calculated in this work (solid lines), at temperatures of 200K (black) and 300K (blue). In the zoomed region, we also added (blue error bars) the experimental CIAs presented in Section 2. More detailed comparisons of the calculated CIAs with the room temperature measurements are provided in Appendix A (Fig. 5 and 8).

## 4. Effect of revised CIAs on the surface temperature of early Mars

We used the LMD Generic numerical climate model in (1D time-marching radiative-convective mode, as in Turbet & Tran, 2017) to perform numerical simulations of $CO_2$-dominated atmospheres (complemented with $H_2$, $CH_4$ and $H_2O$), under Noachian Mars insolation conditions (75% of present-day Mars insolation). The LMD Generic model has previously been used to simulate the surface and atmosphere of early Mars in 3D mode (Forget et al. 2013, Wordsworth et al., 2013, 2015, Turbet et al., 2017, 2020, Turbet & Forget, 2019) and in 1D mode (Wordsworth et al., 2010, Turbet & Tran, 2017, Turbet et al., 2020).

Following Wordsworth et al. (2017), we performed numerical climate calculations for atmospheres of various surface pressures (0.5-2bar), as well as various hydrogen (0.005-0.1) and methane (0.005-0.1) mixing ratios, now including the $CO_2$-$H_2$ and $CO_2$-$CH_4$ collision-induced absorption parameterizations presented in Section 3. The calculations were performed over 30 atmospheric layers from the surface up to a minimum atmospheric pressure of ~10 Pa. Following Wordsworth et al. (2017), $CO_2$, $H_2$ and $CH_4$ mixing ratios were assumed to be constant in the 30 atmospheric layers. $H_2O$ mixing ratio was calculated at each layer assuming a constant tropospheric relative humidity of 0.8. Figure 3 compares our results with those of Wordsworth et al. (2017), for both $CO_2$+$H_2$ (upper panel) and $CO_2$+$CH_4$ (lower panel) atmospheres, and a total atmospheric surface pressure of 2 bar, which corresponds roughly to the upper estimate during the Noachian epoch (Kite, 2019) based on small crater statistics records (Kite et al., 2014) and maximum escape rates extrapolated from MAVEN observations (Jakosky et al., 2018). A more detailed Figure (i.e. comparing results across several surface pressures) is also proposed in Appendix B.

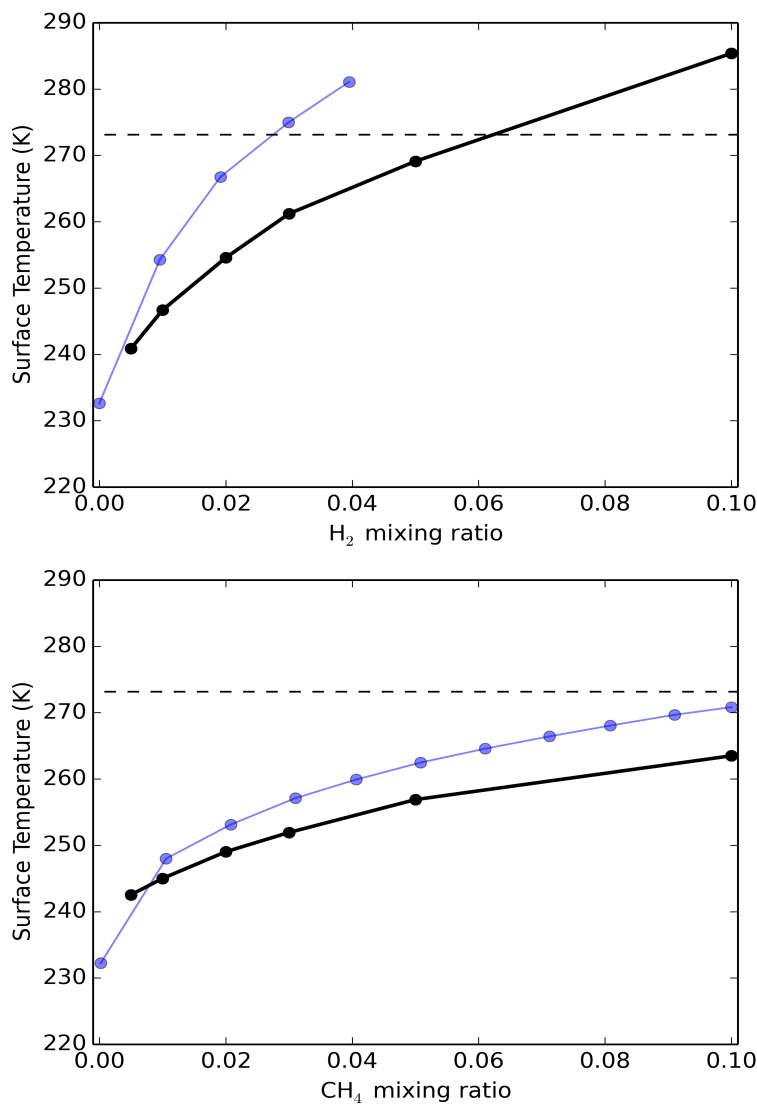

*Figure 3:* This figure shows calculated surface temperature for a 2 bar atmosphere under early Mars condition (75% of present-day Mars insolation; surface albedo taken equal to 0.2), assuming a $CO_2$+$H_2$ atmosphere (upper panel) and a $CO_2$+$CH_4$ atmosphere (lower panel). Black lines show the calculated surface temperatures as a function of $H_2$ and $CH_4$ mixing ratios, using the CIA parameterizations presented in Section 3. For comparison, we added blue lines, which are the calculated surface temperatures of $CO_2$-$H_2$ and $CO_2$-$CH_4$ atmospheres, respectively, taken from Wordsworth et al. (2017). The horizontal dashed black lines indicate the melting temperature of water, i.e. 273.15K. Note that $CH_4$ short-wave absorption is taken into account in our radiative transfer calculations.

Figure 3 (upper panel) shows that in a $CO_2$-dominated 2 bar atmosphere, a ~6% $H_2$ mixing ratio is required to warm the surface above the melting point of water. This mixing ratio estimate is approximately twice that calculated with the $CO_2$-$H_2$ CIA parameterization of Wordsworth et al. 2017. For a $CO_2$+$CH_4$ atmosphere (Fig. 3, lower panel), we calculated that the maximum surface temperature that can be reached is ~263K, about 8K lower than calculated with the $CO_2$-$CH_4$ CIA parameterization of Wordsworth et al. 2017. This is maximal in the sense that for $CH_4$/$CO_2$ mixing ratios higher than roughly 0.1, it is expected that reflective, anti-greenhouse haze would be produced (Trainer et al. 2006). This result confirms that $CO_2$ and $CH_4$ alone (possibly complemented with $N_2$ and $H_2O$) are unable to warm the surface of early Mars above the melting point of water. This makes $H_2$ the most efficient known and plausible gas to warm the surface of Mars above the melting point of water.

Finally, we performed several sensitivity tests to better understand the influence of the wavelength coverage and of the temperature dependence of the $CO_2$-$H_2$ CIA on the calculated surface temperatures. In the first, we artificially suppressed the temperature dependence of the $CO_2$+$H_2$ CIA and fixed the CIA to its 300K value. Doing this led, for a 1bar $CO_2$ atmosphere with 10% $H_2$, to a 12K decrease of the surface temperature, highlighting the importance of knowing accurately the temperature dependence of the CIA. In a second test, we artificially cut the $CO_2$+$H_2$ CIA wavelength coverage beyond 550cm$^{-1}$ (i.e. we retained only the 0-550cm$^{-1}$ part of the CIA). Doing this led, for a 1bar $CO_2$ atmosphere with 10% $H_2$, to an 8K decrease of the surface temperature. Again, this demonstrates that it is not only important to quantify the CIA between 0 and 550 cm$^{-1}$ (because there is a major $CO_2$ atmospheric window around 350cm$^{-1}$; corresponding to the low-frequency wing of the $\nu_2$ band of $CO_2$) but also between 800 and 1200cm$^{-1}$ (corresponding to the high-frequency wing of the $\nu_2$ band of $CO_2$).

## 5. Conclusions and future work

In this paper, we performed a series of experiments and calculations to better constrain the $CO_2$+$CH_4$ and $CO_2$+$H_2$ collision-induced absorptions (CIAs) as well as their effect on the prediction of early Mars surface temperature. First, we carried a new set of experimental measurements (using the AILES line of the SOLEIL synchrotron) of both $CO_2$+$CH_4$ and $CO_2$+$H_2$ CIAs. These measurements confirm the previous results of Turbet et al. (2019) while significantly reducing the uncertainties on the CIA determinations. Secondly, we fitted a semi-empirical model on these CIAs measurements, allowing us to compute $CO_2$+$CH_4$ and $CO_2$+$H_2$ CIAs across a broad spectral domain (0-1500cm$^{-1}$) and for a wide range of temperatures (100-600K). These CIAs are provided in the Supplementary Materials of this paper. Last, we performed 1-D numerical climate calculations (using the LMD Generic Model) to compute the surface temperature expected on the surface of early Mars for several $CO_2$, $CH_4$ and $H_2$ partial pressures, taking into account the radiative effect of the revised CIAs. We found that the concentration of $H_2$ required to warm the surface of early Mars above the melting point of water is 2-3 times higher than previously reported in Wordsworth et al. (2017). However, we showed that a $CO_2$+$CH_4$ atmosphere (assuming a total $CO_2$ surface pressure lower or equal to 2bar, and a $CH_4$/$CO_2$ ratio lower or equal to 0.1) should be unable to warm the surface of early Mars above 273K. This demonstrates that $H_2$ (in a $CO_2$-dominated atmosphere) remains the only known and plausible gas that may possibly raise the annual mean surface temperature of early Mars over the melting temperature of water.

In the future, there are several ways to make progresses on our understanding of the reducing atmosphere scenario which we now – according to our results – suggest renaming the "$H_2$ scenario" or "$H_2$ hypothesis". First, independent measurements of $CO_2$+$H_2$ and/or $CO_2$+$CH_4$ CIAs in a range of wavenumbers and temperatures as large as possible (e.g. see ongoing efforts in Godin et al.,

2019) would be helpful to validate the experimental results presented in our work and in the work of Turbet et al. 2019. Specifically, measurements of the CIAs in the high frequency part of the $\nu_2$ $CO_2$ band (i.e. below 15μm band) and/or at very low or high temperatures would be useful to validate and constrain the semi-empirical model presented in our work.

Then, sophisticated models such as anisotropic fully quantum calculations (Karman et al., 2015) could be developed to produce reliable CIA spectra. Preliminary calculations show however that this approach will be computationally very expensive. Another strategy could be to perform classical molecular dynamics simulations (CMDS), in the same vein than Hartmann et al. (2011), taking possibly into account anisotropic effects as they were shown to be important in our analysis.

Finally, the "$H_2$ scenario" could be evaluated with 3-dimensions Global Climate Models (GCMs). The sophisticated climate models can be used not only to better evaluate the necessary concentration of $H_2$ required to warm the surface of Mars, but can also be used to get insights on the nature of the hydrological cycle expected in such environments. For instance, GCMs can be used to predict erosion spatial patterns which can then be compared with observations of the ancient surfaces of Mars. There are currently many efforts being made to move toward this direction with the 3-D LMD Generic GCM (Turbet, 2018, Turbet & Forget 2019), the NASA AMES GCM (Steakley et al., 2019) and the DRAMATIC GCM (Kamada et al. 2020).

**Acknowledgments**

This project has received funding from the European Union's Horizon 2020 research and innovation program under the Marie Sklodowska-Curie Grant Agreement No. 832738/ESCAPE. M.T. thanks the Gruber Foundation for its support to this research. This work has been carried out within the framework of the National Centre of Competence in Research PlanetS supported by the Swiss National Science Foundation. M.T. acknowledges the financial support of the SNSF. M.T. is grateful for the computing resources on OCCIGEN (CINES, French National HPC).

The authors warmly thank Jean-Michel Hartmann, Ha Tran, and Olivier Pirali for their support and involvement on the CIA measurements experimental campaigns, as well as for their feedbacks on the manuscript. The authors also thank Jean-Michel Hartmann for his strong involvement in the making and structuring of this project. The authors thank Robin Wordsworth for providing $CO_2$, $CH_4$, and $H_2O$ allowed absorption spectra. The authors thank Edwin Kite for pointing a typo in the preprint version of the manuscript.

**References**

Bastien, L.A., Phillip, N.P., Brown, N.J. 2010; Intermolecular potential parameters and combining rules determined from viscosity data. Int. J. Chem. Kinet. 42, 713-723.

Batalha, N., Domagal-Goldman, S. D., Ramirez, R., Kasting, J. F., 2015. Testing the early Mars $H_2$-$CO_2$ greenhouse hypothesis with a 1-D photochemical model. Icarus, Volume 258, p. 337-349.

Boissoles, J., Tipping, R.H., Boulet C. 1994. Theoretical study of the collision induced fundamental absorption spectra of $N_2$-$N_2$ pairs for temperatures between 77 and 297K. J. Quant. Spectrosc. Radiat. Transfer 51, 615-627.

Borysow, A., Moraldi, M. 1992. Effects of anisotropy interaction on collision induced absorption by pairs of linear molecules. Phys. Rev. Letters 68, 3686-3689.

Borysow, A., Tang, C. 1993. Far- infrared spectra of $N_2$-$CH_4$ pairs for modeling of Titan's atmosphere. Icarus 105, 175-183.

Buser, M., Frommhold, L., Moraldi,M., Champagne, M.H., Hunt, K.L.C. 2004. Far-infrared absorption by collisionally interacting nitrogen and methane molecules. J. Chem. Phys. 121, 2617-2621.

Chassefière, E., Lasue, J., Langlais, B., Quesnel, Y, 2016. Early Mars serpentinization-derived $CH_4$ reservoirs, $H_2$-induced warming and paleopressure evolution. Meteoritics & Planetary Science, Volume 51, Issue 11, pp. 2234-2245.

Cherepanov, V.N., Kalugina, Y.N.,Buldakov, M.A., 2016. Interaction-Induced Electric Properties of Van Der Waals Complexes, Springer (New-York).

Forget, F., Wordsworth, R., Millour, E., Madeleine, J. -B., Kerber, L., Leconte, J., Marcq, E., Haberle, R. M., 2013. 3D modelling of the early martian climate under a denser $CO_2$ atmosphere: Temperatures and $CO_2$ ice clouds. Icarus, Volume 222, Issue 1, p. 81-99.

Godin, P.J., Ramirez, R.M., Campbell, C., Wizenberg, T., Nguyen, T.G. , Strong, K., Moores, J.E. 2019. Collision-Induced Absorption of $CH_4$-$CO_2$ and $H_2$-$CO_2$ Complexes and Their Effect on the Ancient Martian Atmosphere. ESSOAr | https:/doi.org/10.1002/essoar.10501475.1

Gruszka, M., Borysow, A., 1996. New analysis of the spectral moments of collision induced absorption in gaseous $N_2$ and $CO_2$. Mol. Phys. 88, 1173-1185.

Haberle, R. M., Catling, D. C., Carr, M. H., Zahnle, K. J., 2017. The Early Mars Climate System. The atmosphere and climate of Mars, Edited by R.M. Haberle et al. Cambridge University Press, 2017, p. 497-525.

Haberle, R.M., Zahnle, K., Barlow, N.G., Steakley, K.E., 2019. Impact Degassing of $H_2$ on Early Mars and Its Effect on the Climate System. Geophysical Research Letters, in press.

Hartmann, J.-M., Boulet, C., Jacquemart, D., 2011. Molecular dynamics simulations for $CO_2$ spectra. II. The far infrared collision-induced absorption band. Journal of Chemical Physics, Volume 134, Issue 9, pp. 094316-094316-9.

Hellmann, R., Bich, E., Vesovic, V., 2016. Cross second virial coefficients and dilute gas transport properties of the ($CH_4$+$CO_2$), ($CH_4$+$H_2S$) and ($H_2S$+$CO_2$) systems from accurate intermolecular potential energy surfaces. J. Chem. Thermodynamics 102, 429-441.

Hunt, J.L., Poll, J.D., 1978. Lineshape analysis of collision induced spectra of gases. Can. J. Phys. 56, 950-961.

Jakosky, B. M., Brain, D., Chaffin, M., Curry, S., Deighan, J., Grebowsky, J., Halekas, J., Leblanc, F., Lillis, R., Luhmann, J. G., Andersson, L., Andre, N., Andrews, D., Baird, D., Baker, D., Bell, J., Benna, M., Bhattacharyya, D., Bougher, S., Bowers, C. Chamberlin, P., Chaufray, J. -Y., Clarke, J., Collinson, G., Combi, M., Connerney, J., Connour, K., Correira, J., Crabb, K., Crary, F., Cravens, T., Crismani, M., Delory, G., Dewey, R., DiBraccio, G., Dong, C., Dong, Y., Dunn, P., Egan, H., Elrod, M., England, S., Eparvier, F., Ergun, R., Eriksson, A., Esman, T., Espley, J., Evans, S., Fallows, K., Fang, X., Fillingim, M., Flynn, C., Fogle, A., Fowler, C., Fox, J., Fujimoto, M., Garnier, P., Girazian, Z., Groeller, H., Gruesbeck, J., Hamil, O., Hanley, K. G., Hara, T., Harada, Y., Hermann, J., Holmberg, M., Holsclaw, G., Houston, S., Inui, S., Jain, S., Jolitz, R., Kotova, A.,

Kuroda, T., Larson, D., Lee, Y., Lee, C., Lefevre, F., Lentz, C., Lo, D., Lugo, R., Ma, Y. -J., Mahaffy, P., Marquette, M. L., Matsumoto, Y., Mayyasi, M., Mazelle, C., McClintock, W., McFadden, J., Medvedev, A., Mendillo, M., Meziane, K., Milby, Z., Mitchell, D., Modolo, R., Montmessin, F., Nagy, A., Nakagawa, H., Narvaez, C., Olsen, K., Pawlowski, D., Peterson, W., Rahmati, A., Roeten, K., Romanelli, N., Ruhunusiri, S., Russell, C., Sakai, S., Schneider, N., Seki, K., Sharrar, R., Shaver, S., Siskind, D. E., Slipski, M., Soobiah, Y., Steckiewicz, M., Stevens, M. H., Stewart, I., Stiepen, A., Stone, S., Tenishev, V., Terada, N., Terada, K., Thiemann, E., Tolson, R., Toth, G., Trovato, J., Vogt, M., Weber, T., Withers, P., Xu, S., Yelle, R., Yiğit, E., Zurek, R., 2018. Loss of the Martian atmosphere to space: Present-day loss rates determined from MAVEN observations and integrated loss through time. Icarus, Volume 315, p. 146-157.

Kamada, A., Kuroda, T., Kasaba, Y., Terada, N., Nakagawa, H., Toriumi, K., 2020. A coupled atmosphere–hydrosphere global climate model of early Mars: A 'cool and wet' scenario for the formation of water channels. Icarus, Volume 338, 1 March 2020, 113567.

Karman, T., Miliordos, E., Hunt, K.L.C., Groenenboom, C., 2015. Quantum mechanical calculation of the collision-induced absorption spectra of $N_2$-$N_2$ with anisotropic interaction. J. Chem. Phys. 142, 084306.

Karman, T., Gordon, I. E., van der Avoird, A., Baranov, Y. I., Boulet, C., Drouin, B. J., Groenenboom, G. C., Gustafsson, M., Hartmann, J.-M., Kurucz, R. L., Rothman, L. S., Sun, K., Sung, K., Thalman, R., Tran, H., Wishnow, E. H., Wordsworth, R., Vigasin, A. A., Volkamer, R., van der Zande, W. J., 2019. Update of the HITRAN collision-induced absorption section. Icarus, Volume 328, p.160-175.

Kite, E. S., Williams, J.-P., Lucas, A., Aharonson, O., 2014. Low palaeopressure of the martian atmosphere estimated from the size distribution of ancient craters. Nature Geoscience, Volume 7, Issue 5, pp. 335-339.

Kite, E. S., 2019. Geologic Constraints on Early Mars Climate. Space Science Reviews, Volume 215, Issue 1, article id. 10, 47 pp.

Kite, E.S., Mischna, M., Gao, P., Yung, Y.L., Turbet, M., 2020. 'Methane release on Early Mars by atmospheric collapse and atmospheric reinflation', Planetary and Space Science, in press.

Leforestier, C., Tipping, R.H., Ma, Q., 2010. Temperature dependences of mechanisms responsible for the water-vapor continuum absorption. II. Dimers and collision-induced absorption. J. Chem. Phys. 132, 164302.

Li, X., Champagne, H., Hunt, K.L.C. 1998. Long-range, collision induced dipoles of $T_d - D_{\infty h}$ molecule pairs: Theory and numerical results for CH4 or CF4 interacting with H2, N2, CO2, or CS2. J. Chem. Phys. 109, 8416-8425.

Li, H., Roy, P.N., Le Roy, R.J., 2010. Analytic Morse/long range potential energy surfaces and predicted infrared spectra for $CO_2$-$H_2$. J. Chem. Phys. 132, 214309.

Poll, J.D., Hunt, J.L., 1976. On the moments of the pressure induced spectra of gases. Can. J. Phys. 54, 461-470.

Ramirez, R. M., Kopparapu, R., Zugger, M. E., Robinson, T. D., Freedman, R., Kasting, J. F., 2014. Warming early Mars with $CO_2$ and $H_2$. Nature Geoscience, Volume 7, Issue 1, pp. 59-63.

Ramirez, R. M., 2017. A warmer and wetter solution for early Mars and the challenges with transient warming. Icarus, Volume 297, p. 71-82.

Ramirez, R. M & Craddock, R. A, 2018. The geological and climatological case for a warmer and wetter early Mars. Nature Geoscience, Volume 11, Issue 4, p.230-237.

Richard, C., Gordon, I. E., Rothman, L. S., Abel, M., Frommhold, L., Gustafsson, M., Hartmann, J.-M., Hermans, C., Lafferty, W. J., Orton, G. S., Smith, K. M., Tran, H., 2012. New section of the HITRAN database: Collision-induced absorption (CIA). Journal of Quantitative Spectroscopy and Radiative Transfer, Volume 113, Issue 11, p.1276-1285.

Steakley, K., Kahre, M., Haberle, R., Zahnle, K., 2019. Exploring post-impact reducing greenhouse climates for early Mars with the NASA Ames Mars Global Climate Model. EPSC-DPS Joint Meeting 2019, held 15-20 September 2019 in Geneva, Switzerland, id. EPSC-DPS2019-509.

Tarnas, J. D., Mustard, J. F., Sherwood Lollar, B., Bramble, M. S., Cannon, K. M., Palumbo, A. M., Plesa, A.-C, 2018. Radiolytic $H_2$ production on Noachian Mars: Implications for habitability and atmospheric warming. Earth and Planetary Science Letters, Volume 502, p.133-145.

Trainer, M. G., Pavlov, A. A., Dewitt, H. L., Jimenez, J. L., McKay, C. P., Toon, O. B., Tolbert, M. A., 2006. Inaugural Article: Organic haze on Titan and the early Earth. Proceedings of the National Academy of Sciences, vol. 103, issue 48, pp. 18035-18042.

Turbet, M., Forget, F., Head, J. W., Wordsworth, R., 2017. 3D modelling of the climatic impact of outflow channel formation events on early Mars. Icarus, Volume 288, p. 10-36.

Turbet, M. & Tran, H., 2017. Comment on "Radiative Transfer in $CO_2$-Rich Atmospheres: 1. Collisional Line Mixing Implies a Colder Early Mars". Journal of Geophysical Research: Planets, Volume 122, Issue 11, pp. 2362-2365.

Turbet, M., 2018. Habitability of Planets Using Numerical Climate Models. Application to Extrasolar Planets and Early Mars. Theses. Sorbonne Université/Université Pierre et Marie Curie - Paris VI.

Turbet, M., Tran, H., Pirali, O., Forget, F., Boulet, C., Hartmann, J-M., 2019. Far infrared measurements of absorptions by $CH_4 + CO_2$ and $H_2 + CO_2$ mixtures and implications for greenhouse warming on early Mars. Icarus 321, 189-199.

Turbet, M., Forget, F., 2019. The paradoxes of the Late Hesperian Mars ocean. Scientific Reports, Volume 9, id. 5717.

Turbet, M., Gillmann, C., Forget, F., Baudin, B., Palumbo, A., Head, J., Karatekin, O., 2020. The environmental effects of very large bolide impacts on early Mars explored with a hierarchy of numerical models. Icarus, Volume 335, article id. 113419.

Wordsworth, R., Forget, F., Eymet, V., 2010. Infrared collision-induced and far-line absorption in dense $CO_2$ atmospheres. Icarus, Volume 210, Issue 2, p. 992-997.

Wordsworth, R., Forget, F., Millour, E., Head, J. W., Madeleine, J.-B., Charnay, B., 2013. Global modelling of the early martian climate under a denser $CO_2$ atmosphere: Water cycle and ice evolution. Icarus, Volume 222, Issue 1, p. 1-19.

Wordsworth, R. D., Kerber, L., Pierrehumbert, R. T., Forget, F., Head, J. W., 2015. Comparison of "warm and wet" and "cold and icy" scenarios for early Mars in a 3-D climate model. Journal of Geophysical Research: Planets, Volume 120, Issue 6, pp. 1201-1219.

Wordsworth, R. D., 2016. The Climate of Early Mars. Annual Review of Earth and Planetary Sciences, vol. 44, p.381-408.

Wordsworth, R, Kalugina, Y, Lokshtanov, S, Vigasin, A, Ehlmann, B, Head, J, Sanders, C, Wang, H, 2017. Transient reducing greenhouse warming on early Mars. Geophys. Res. Lett. 44, 665-671.

Zvereva-Loete,N., Kalugina, Y.N., Boudon, V., Buldakov, M.A., Cherepanov, V.N. 2010. Dipole moment surface of the van der Waals complex CH4-N2. J. Chem. Phys. 133, 184302.

# Appendix A - Semi-empirical CIA model

In this Appendix, we describe in details how we constructed our semi-empirical model of $CO_2$-$H_2$ and $CO_2$-$CH_4$ CIAs.

## A.1. $CO_2$-$H_2$

We first start this Appendix by introducing the various terms and approximations necessary for the readers to understand the content and philosophy of our semi-empirical models of $CO_2$+$H_2$ (and $CO_2$+$CH_4$) collision-induced absorptions.

*The induced dipole (ID):* As discussed by Poll & Hunt (1976), the spherical components of the ID in a space fixed system, for a pair of linear molecules can be written in terms of coupled spherical harmonics specified by four indices $\lambda_1, \lambda_2, \Lambda \wedge L$ satisfying some general constraints. Following a preliminary study of Cherepanov et al. 2016 (see p.28), showing that the short range components of the ID are negligible for distances R (between the centers of mass) greater than about $7a_0$, we have limited the ID to its long range components. We have taken into account all the components as given in the appendix of Hartmann et al. (2011). Note that all the molecular parameters appearing in these expressions are well known and may be easily found in the literature (Li et al., 1998, Hartmann et al., 2011, and references therein).

*The intermolecular potential:* Reliable anisotropic potentials are now available for the $CO_2$+$H_2$ pair. In the present work, we have chosen the potential proposed by Li et al. (2010), which gave good predictions of the IR spectra of the complex. Its isotropic part can be easily determined through an "ad-hoc" averaging over the relative orientations.

*The classical isotropic approximation (IA):* In a classical approach, the pair distribution function is given by (Borysow & Moraldi, 1992):

$$g(R,\Omega) = exp\left(-V(R,\Omega)/k_B T\right) \quad \text{(A-1)}$$

where $V(R,\Omega)$ is the intermolecular potential, with $R$ the intermolecular distance and $\Omega$ indicating the relative orientation of the two molecules. In the IA, one assumes that the potential may be limited to its isotropic part $V_0(R)$.

*The absorption coefficient:* As the theory has been extensively presented in detail elsewhere (see for instance Boissoles et al., 1994), only a brief review is given here. Within the IA, the CIA coefficient (in cm$^{-1}$ amagat$^{-2}$) can be written as a sum of rotational components, each having a translational profile:

$$\alpha_{iso}(\omega) = \frac{4\pi^2}{3\hbar c} n_0^2 \omega \left(1 - e^{-\hbar\omega/k_B T}\right) \sum_n \Gamma_n(T) G_n(\omega - \omega_n, T) \quad \text{(A-2)}$$

Where $n_0$ is the number density at normal temperature and pressure. Each component $n$ is specified by a set of quantum numbers, $n = \{j_1, j_2, j'_1, j'_2, \lambda_1, \lambda_2, \Lambda \wedge L\}$, in which $j_i$ designates the rotational quantum number of $CO_2$ (i=1), and $H_2$ (i=2); a prime indicates final states. $\Gamma_n(T)$ represents the strength of the rotational component while $\omega_n = \omega_{j_1 j'_1} + \omega_{j_2 j'_2}$ is its central frequency. Their expressions may be found in Boissoles et al. (1994).

*The translational profile:* In a fully quantum but isotropic approach, the product $\Gamma_n(T)G_n(\omega,T)$ can be directly calculated from the isotropic potential by solving a Schrödinger equation (see Karman et al., 2015). Here we preferred to use a more empirical lineshape. There have been many models proposed in the past. In the present work, we have chosen the so-called K2 lineshape to represent the translational profile (Hunt and Poll, 1978):

$$G_n(\omega, T) = \frac{2}{3\pi\eta_n} \frac{1}{1+e^{-\hbar\omega/k_B T}} \left(\frac{\omega}{\eta_n}\right)^2 K_2\left(\frac{\omega}{\eta_n}\right) \quad \text{(A-3)}$$

where $K_2$ is a modified Bessel function of the second kind. $\eta_n$ is a parameter of the same dimension as $\omega$. The main advantage of the K2 lineshape is that $\eta_n$ can be determined, for each temperature, from the only knowledge of the isotropic potential and the corresponding spectral moments. Details on how this can be done are available, for instance, in Leforestier et al. (2010).

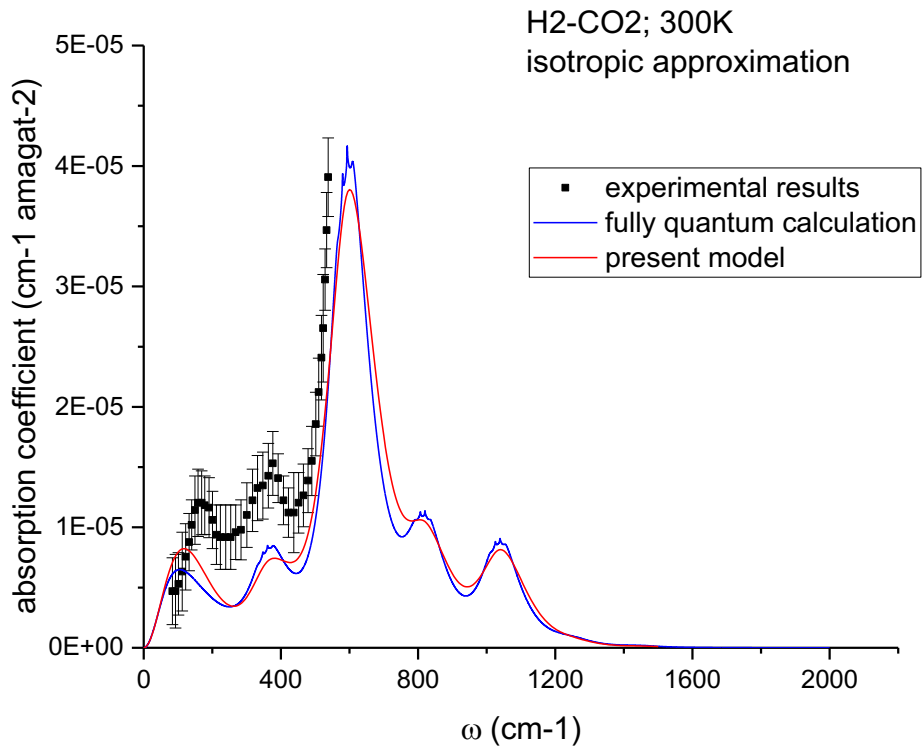

*Figure 4:* Comparison between the experimental results and the theoretical calculations of the $CO_2$-$H_2$ CIA within the isotropic approximation.

*Results in the isotropic approximation:* Once the rotational strengths and translational profiles are available, the CIA coefficient at T=296K can be calculated and compared with the experimental data. As shown in Fig. 4, this calculation does not provide sufficient spectral intensity to explain the measured absorption.
Of course, the present model is subject to many uncertainties. It was, therefore, of some interest to compare its predictions with a more "exact" calculation in order to determine its failures. As mentioned above, quantum calculations are now feasible for $CO_2$-$H_2$, within the isotropic and coupled states approximations, avoiding therefore the use of a classical pair distribution function (appearing in the rotational strengths) and empirical translational profiles. In Karman et al. (2015), this formalism has been applied to the $N_2$-$N_2$ CIA, leading to a good agreement with the measured data in a wide range of temperatures. By using the same isotropic potential and the same ID components as in our semi-empirical model, the "isotropic quantum" spectrum has been calculated at 296K and is compared with the more simple semi-empirical model results in Fig. 4. There is a very good agreement between the two theoretical calculations, thus validating the use of the semi-empirical approach, which is much easier to handle. However, both calculations disagree with the measured absorption, indicating that anisotropic effects may be important to take into account here.

*Failure of the isotropic approximation:* It appears more clearly that the validity of the isotropic approximation needs to be analyzed. A first indication on its accuracy can be obtained by examining the anisotropic potential of Li et al. (2010) which exhibits a strong dependence on the mutual orientation of the two colliding molecules. Based on a spectral moments analysis, the inadequacy of the IA can then be easily demonstrated. Gruszka & Borysow (1996) have developed a formalism for the calculation of the zeroth order spectral moment and of the integrated intensity of the band, which fully accounts for the anisotropy of the potential. We have applied this formalism to the $CO_2$-$H_2$ pair. The results, for some selected temperatures are given in Tables 1 and 2, and compared with those based on the IA. Even if spectral moments contain less information than spectral densities, this comparison provides a significant test of the inadequacy of the IA. Before going on, note that our results for the "anisotropic" $0^{th}$ order spectral moments are in reasonable agreement with those calculated by Wordsworth et al. (2017), which are equal to 1.6-1.8×$10^{-4}$ cm$^{-1}$ between 200-300K, respectively (see Fig. 1 in the Supporting Information of Wordsworth et al., 2017).

Table 1: Integrated intensity S (in $10^{-2}$ cm$^{-2}$ amagat$^{-2}$). As a reminder, $S = \int_0^\infty d\omega \, \alpha(\omega)$ is the integral of the CIA binary coefficient $\alpha(\omega)$ over wavenumber $\omega$.

| T(K) | $S_{iso}(T)$ | $S_{aniso}(T)$ | ratio |
|---|---|---|---|
| 100 | 1.38 | 4.37 | 3.17 |
| 200 | 1.27 | 2.54 | 2 |
| 300 | 1.3 | 2.4 | 1.85 |
| 500 | 1.42 | 2.46 | 1.73 |

Table 2: Zeroth order moment (in $10^{-4}$ cm$^{-1}$ amagat$^{-2}$).
As a reminder, $M^{(0)} = \int_0^\infty d\omega \, \frac{\alpha(\omega)}{\omega} \, \frac{1}{\tanh(\frac{\hbar\omega}{2k_BT})}$ is the zero$^{th}$ order moment of the CIA binary coefficient $\alpha(\omega)$.

| T(K) | $M^{(0)}_{iso}(T)$ | $M^{(0)}_{aniso}(T)$ | ratio |
|---|---|---|---|
| 100 | 1.15 | 4.03 | 3.17 |
| 200 | 1.07 | 2.16 | 2.02 |
| 300 | 1.09 | 2.01 | 1.84 |
| 500 | 1.18 | 2.05 | 1.74 |

*Empirical correction of the isotropic spectral densities:* As mentioned above, accounting for the anisotropy of the intermolecular potential remains a challenge for fully quantum calculations with a heavy collision partner like $CO_2$. Here we rather built on our semi-empirical model that we modified to account for anisotropic effects using the following procedure:

i   We assumed that the area-normalized absorption coefficient (i.e. the spectral shape) can be reasonably calculated within the IA, from eq. (A-2).
ii  We multiplied the normalized profile by the "true" anisotropic integrated intensity (provided in Table 2):

$$\alpha_{aniso}(\omega, T) = \frac{S_{aniso}(T)}{S_{iso}(T)} \alpha_{iso}(\omega, T) \qquad (A-4)$$

Figure 5 shows that this procedure leads to a theoretical CIA coefficient that is in good agreement with the available experimental data at 296K. We then computed the $CO_2$-$H_2$ CIA in a wide range of temperatures using the temperature dependencies of the parameters of our models. Results are provided in Figure 6. Note that, because the calculated zeroth order spectral moments are quasi-independent of temperature within the 200-300K range (which is the typical range of temperatures expected in the atmosphere of early Mars) as seen in Table 2, our model should be reliable for early Mars applications.

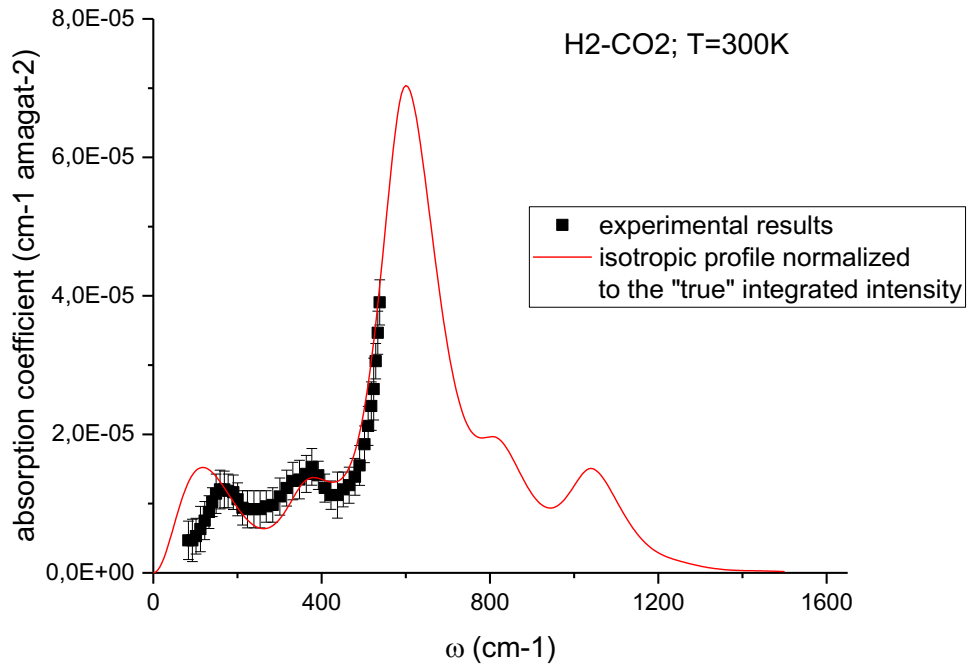

*Figure 5: Comparison between the experimental results (presented in Section 2) and the renormalized, anisotropic $CO_2+H_2$ collision-induced absorption coefficient.*

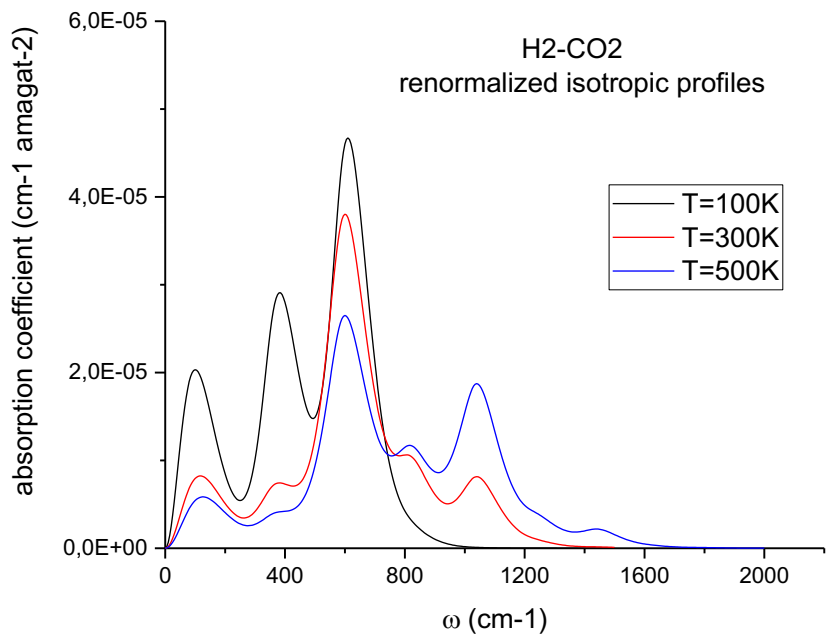

*Figure 6: Temperature dependence of the renormalized, anisotropic $CO_2+H_2$ collision-induced absorption coefficient.*

## A.2. $CO_2$-$CH_4$

Before considering the $CO_2$+$CH_4$ pair, let us recall the state of the art for a very similar system, $N_2$-$CH_4$, for which many comparisons between theory and measurements have been made. The most recent analysis has been proposed by Buser et al. (2004). Their main approximations were:

i  The isotropic approximation (IA), but using an "effective" isotropic potential, i.e. determined from fitting various scattering data. This effective isotropic potential may include part of the effects of the true anisotropic potential.

ii  The induced dipole (ID) limited to its long-range components (calculated with well-known molecular parameters requiring no adjustment).

For wavenumbers $\geq 175 cm^{-1}$, theory and measurements diverge more and more, reaching one order of magnitude around 1000 $cm^{-1}$ (Buser et al., 2004). To the best of our knowledge, no clear explanation of that discrepancy has been given yet. However, an alternative had been proposed by Borysow & Tang (1993) in order to reproduce all the existing measurements, in a wide range of temperature (70-296K) with the greatest possible accuracy (for support of the analysis of Titan's atmosphere). Their alternative model consists in adding empirical short-range contributions to the $\lambda_1, \lambda_2, L$ ID components 203, 034, 045 and 067 ((1) stands for $N_2$ and (2) for $CH_4$). The corresponding parameters were adjusted through a fitting of all the available experimental spectra. Here we followed a similar procedure for the $CO_2$-$CH_4$ system.

*Isotropic approximation and effective isotropic potential:* A sophisticated anisotropic potential is now available for the $CO_2$+$CH_4$ pair (Hellman et al., 2016). As for $CO_2$-$H_2$ the analysis of the shape of this potential for various mutual orientations of the collisional partners rises some doubt about the validity of the "true" isotropic approximation. By "true" we mean using the "true" isotropic part of the anisotropic potential. For example, from Fig. 2 of Hellman et al. (2016), the distance between the two molecules at which the interaction potential reaches zero may vary from 3Å to about 4.5Å for different mutual orientations. However, less rigorous descriptions of static and dynamic properties for that pair have been made, using an effective isotropic potential adjusted to reproduce the corresponding data. In this work we have used a potential given by Bastien et al. (2010). They fitted viscosity data for binary mixtures and among different solutions proposed a Lennard-Jones 6-12 potential, with the following parameters: $\varepsilon/k_B T = 203 K; \sigma = 3.735 Å$.

*ID components:* In a first step we have considered only the long-range components as given for instance in Borysow & Tang (1993). Here too, the needed molecular parameters are well known and easily found in the literature (Li et al., 1998, Hartmann et al., 2011, Zvereva-Loete et al., 2010, and references therein). In a second step and as stated above, we added an empirical short-range component following the approach of Borysow & Tang (1993).

*Translational shape:* Since the isotropic potential is known, we decided to use K2 lineshapes that were determined following the procedure already outlined for the $CO_2$+$H_2$ pair.

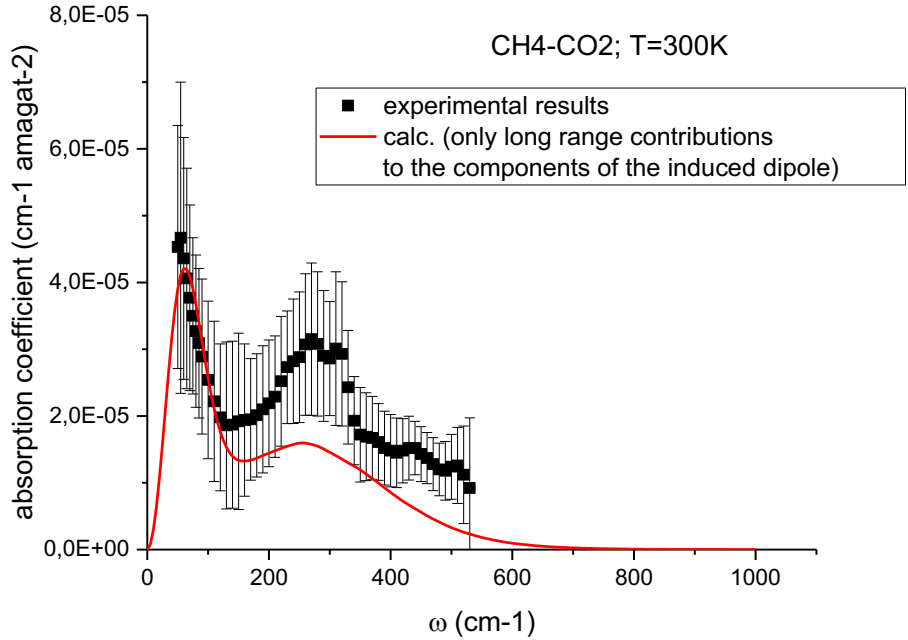

*Figure 7:* Comparison between the experimental data of Turbet et al. (2019) and the prediction of our semi-empirical model for the $CO_2+CH_4$ CIA spectrum. Here the induced dipole (ID) is limited to its long-range components.

*Results:* Figure 7 shows the result of our semi-empirical model when considering only the long-range components of the ID. We observe significant discrepancies between this first model and the experimental data of Turbet et al. (2019) and those presented in the Section 2. The situation is in fact rather similar to that observed for $N_2$-$CH_4$ at the same level of approximation.

Therefore, we have followed the method initially proposed by Borysow & Tang (1993), by adding short range contributions to the same $\lambda_1, \lambda_2, L$ ID components. Fitting the four corresponding parameters requires an adjustment on many experimental profiles measured in a wide range of temperatures, as was done for the $N_2+CH_4$ pair. However, such measurements are not available yet for the $CO_2+CH_4$ pair. We have therefore included the contribution of the short-range components of the $CO_2+CH_4$ pair by using the short-range component of the $N_2+CH_4$ pair multiplied by a constant (a factor 2) adjusted to fit the available experimental data. The result of this second semi-empirical model (including a parameterization of the short-range components) is presented in Figure 8.

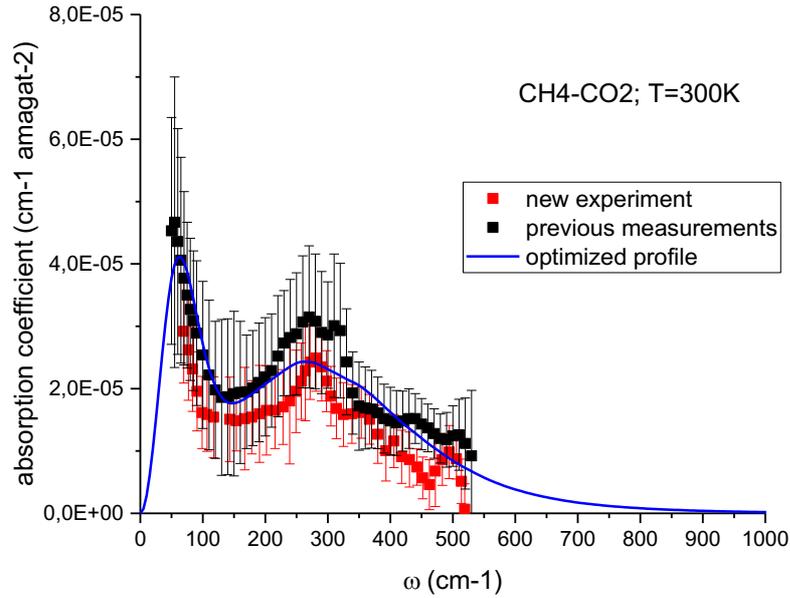

*Figure 8:* *Comparison between the experimental data and the optimized model for the $CO_2+CH_4$ CIA, after introducing short range components to the induced dipole. While "Previous measurements" refer to the work of Turbet et al. 2019, the "new experiment" refer to the work presented in Section 2 of this paper.*

Figure 8 shows that we can obtain a reasonable agreement with the measured absorption at 296K. We then computed the $CO_2$-$CH_4$ CIA in a wide range of temperatures using the temperature dependencies of the parameters of our models. Results are provided in Figure 9.

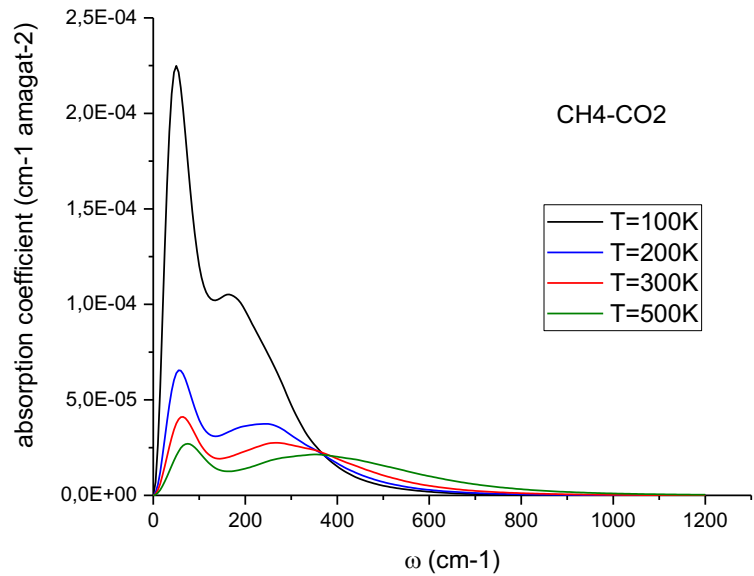

*Figure 9:* *Temperature dependence of the $CO_2$-$CH_4$ collision-induced absorption coefficient, as predicted by the optimized model.*

**Appendix B – Additional Figure on the calculations of early Mars surface temperature**

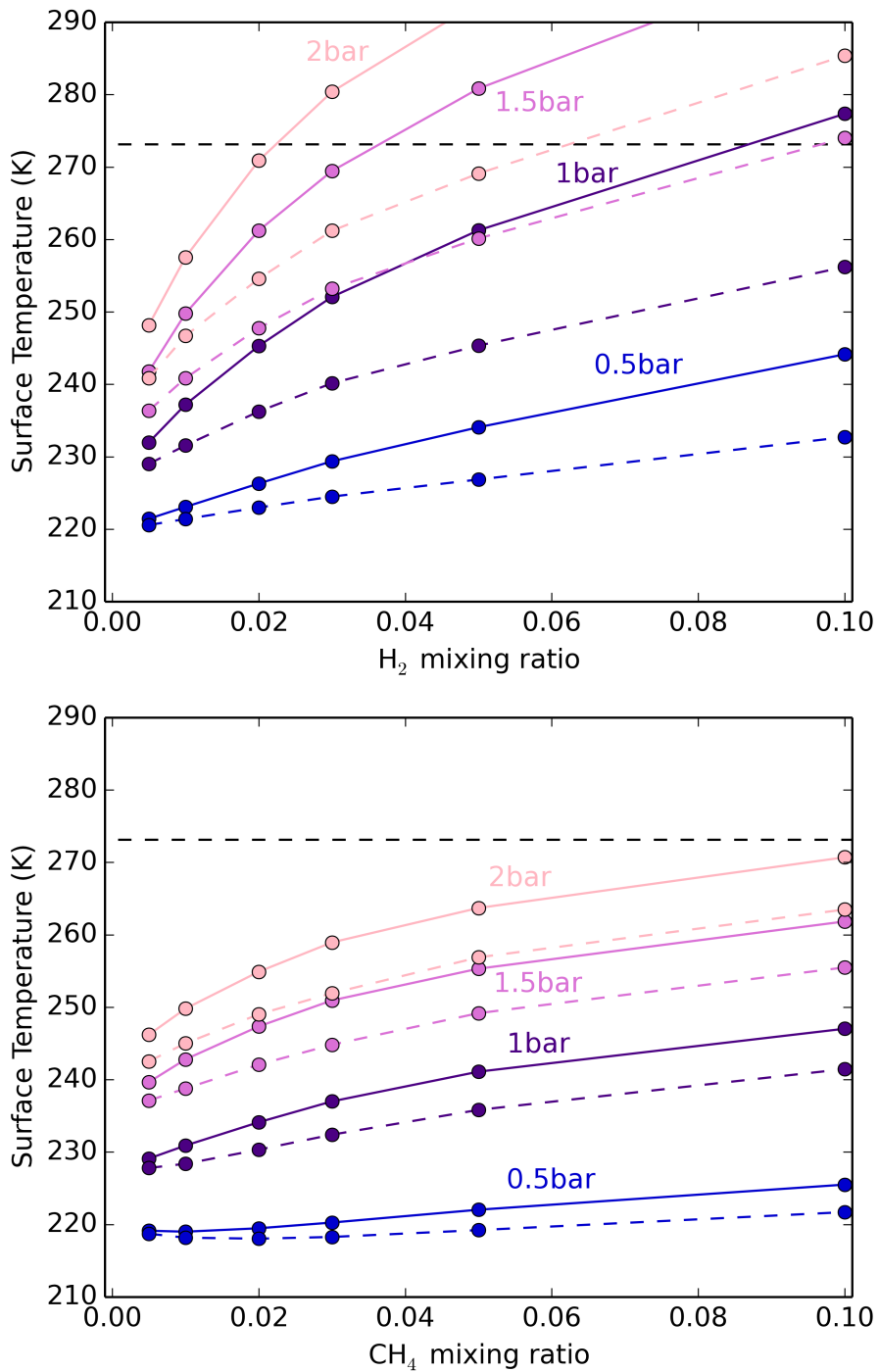

Figure 10: Predicted surface temperature in $CO_2$-dominated atmospheres under early Mars condition (75% of present-day Mars insolation; surface albedo taken equal to 0.2) as a function of $H_2$ and $CH_4$ concentration (upper and lower panels, respectively) and for various surface pressures. Solid lines show results using Wordsworth et al. (2017) CIAs while dashed lines were computed using CIAs presented in the Section 3 of this paper. For reference, the horizontal dashed black lines indicate the melting temperature of water, i.e. 273.15K.